\documentstyle[emulateapj,psfig]{article}

\begin{document}

\title{Afterglow Emission from Highly Collimated Jets with Flat Electron Spectra:
Application to the GRB 010222 Case?}
\author{Z. G. Dai$^1$ and K. S. Cheng$^2$}
\affil{$^1$Department of Astronomy, Nanjing University, Nanjing 210093, China \\ 
$^2$Department of Physics, The University of Hong Kong, Hong Kong, China \\
E-mail: daizigao@public1.ptt.js.cn; hrspksc@hkucc.hku.hk}

\begin{abstract}
We derive light curves of the afterglow emission 
from highly collimated jets if the power-law index 
($p$) of the electron energy distribution is above 1 but below 2.  
We find (1) below the characteristic synchrotron frequency, 
the light curve index depends generally on $p$. (2) As long as the jet 
expansion is spherical, the light curve index above the characteristic 
frequency increases slowly as the spectral index of 
the emission increases. (3) Once the jet enters the spreading phase, 
the high-frequency emission flux decays as  $\propto t^{-(p+6)/4}$ 
rather than $\propto t^{-p}$. All these results differ from those 
in the case of $p>2$. We compare our analytical results with the 
observations on the GRB 010222 afterglow, and conclude that 
the jet model may be unable to explain the observed data. 
Thus, a more promising explanation for this afterglow seems to be 
the expansion of a relativistic fireball or a mildly collimated jet 
in a dense medium.         
\end{abstract}


\section {Introduction}

Gamma-ray burst (GRB) afterglows are believed to be emitted from  
a relativistic shock wave expanding in its surrounding medium via 
synchrotron radiation or inverse Compton scattering (ICS) of accelerated 
electrons in the shocked matter (Piran 1999; van Paradijs, Kouveliotou 
\& Wijers 2000; Cheng \& Lu 2001). To interpret the abundant 
data of afterglows,  the effects of environments such as pre-burst 
stellar winds (Dai \& Lu 1998; M\'esz\'aros, Rees \& Wijers 1998; 
Chevalier \& Li 1999, 2000) and dense media (Dai \& Lu 1999, 2000; 
Wang, Dai \& Lu 2000)  have been discussed. On the other hand, 
jets are of particular interest because they have important implications 
on almost all aspects of the GRB phenomenon, e.g., the total energy that 
is released in an explosion, the event rate, the physical ejection mechanism 
and the afterglow decay rate. The most exciting implication is that 
the transition of a relativistic jet to the spreading phase can result in steepening 
of the afterglow light curve to the flux $\propto t^{-p}$, as analyzed by Rhoads 
(1999) and Sari, Piran \& Halpern (1999). Following the analytical work, 
many numerical calculations have been performed and they are essentially 
consistent with the analytical results (e.g., Panaitescu \& M\'esz\'aros 
1999; Moderski, Sikora \& Bulik 2000; Kumar \& Panaitescu 2000; Panaitescu 
\& Kumar 2001; Huang et al. 2000a, b, c; Wei \& Lu 2000). The jet model 
seems to account well for a few well-observed 
afterglows with light curve breaks, e.g., GRB 990123 (Kulkarni et al. 1999; 
Castro-Tirado et al. 1999; Fruchter et al. 1999), GRB 990510 
(Harrison et al. 1999; Stanek et al. 1999), GRB 991216 
(Halpern et al. 2000), GRB 000301C (Rhoads \& Fruchter 2001; 
Masetti et al. 2000; Jensen et al. 2000; Berger et al. 2000; Sagar et al. 2000), 
GRB 000418 (Berger et al. 2001), and GRB 000926 (Price et al. 2001; 
Harrison et al. 2001; Sagar et al. 2001a; Piro et al. 2001).

GRB 010222 is the latest well-observed burst, whose optical-afterglow 
light curve has an earliest sharp break (Masetti et al. 2001; 
Stanek et al. 2001; Sagar et al. 2001b; Cowsik et al. 2001). A popular 
explanation is that this afterglow might have come from 
a highly collimated jet with a flat-spectrum electron distribution
($1<p<2$). In this Letter we derive light curves of the emission 
when such jets expand in an interstellar medium (ISM) or in a stellar wind, 
and find that the jet model may be inconsistent with the afterglow 
data of GRB 010222.  
  
\section{Light Curves}

Let's assume an {\em adiabatic} relativistic jet with an initial half opening angle 
of $\theta_0$, a laterally-spreading velocity of $c_s$ and a bulk Lorentz 
factor of $\gamma$. This assumption is valid if the energy 
density of the electrons accelerated by a shock, produced by the interaction 
of the jet with its surrounding medium, is a small fraction $\epsilon_e$ of the 
total energy density of the shocked medium or if most of the electrons are 
adiabatic, i.e., their radiative cooling timescale is larger than that of the jet 
expansion (Sari, Piran \& Narayan 1998). The energy density carried by 
magnetic fields is assumed to be another fraction $\epsilon_B$ of the total 
energy density of the shocked medium, and thus the magnetic strength
$B=[32\pi \epsilon_B\gamma^2 n(r)m_pc^2]^{1/2}$, where $m_p$ is the proton 
mass and $n(r)$ is the proton number density of the surrounding medium 
at shock radius $r$. We adopt a power-law density profile: 
$n(r)=Ar^{-s}$, where  $A=n_*\times 1\,{\rm cm}^{-3}$ for the ISM ($s=0$), 
and $A=3\times 10^{35}A_*\,{\rm cm}^{-1}$ for the wind ($s=2$) 
(Chevalier \& Li 1999). 

We consider synchrotron radiation of 
the electrons accelerated by the shock. To calculate the spectrum and 
light curve, one needs to determine three break frequencies: the  
self-absorption frequency ($\nu_a$), the characteristic frequency 
($\nu_m$) and the cooling frequency ($\nu_c$). The latter two frequencies 
can be directly derived from the minimum Lorentz factor $\gamma_m$ and 
the cooling Lorentz factor $\gamma_c$, which appear in the energy 
distribution of cooled electrons. As usual, we adopt a power-law 
injection of electrons with the energy distribution (just behind the shock front)
given by $dn_e/d\gamma_e\propto \gamma_e^{-p}$ for 
$\gamma_m\le\gamma_e\le\gamma_M$, where  $\gamma_M
=[3 e/(\xi\sigma_T B)]^{1/2}$ is the maximum electron Lorentz factor,
which is calculated by assuming that the acceleration time equals the 
synchrotron cooling time. Here $\xi\sim 1$ is the ratio of the acceleration 
time to the gyration time, $e$ is the electron charge, and $\sigma_T$ is the 
Thomson cross section. According to this electron energy distribution and 
the jump conditions for a relativistic shock, the electron number density and 
energy density of the shocked medium can be written as two integrals:
$\int^{\gamma_M}_{\gamma_m}(dn_e/d\gamma_e)d\gamma_e=4\gamma n$ and
$\int^{\gamma_M}_{\gamma_m}(\gamma_em_ec^2)(dn_e/d\gamma_e)d\gamma_e
=4\gamma^2nm_pc^2\epsilon_e$,  where $m_e$ is the electron mass. Such integrals 
combined with the assumption of a flat electron spectrum ($1<p<2$) lead to 
\begin{equation}
\gamma_m=\left[\left(\frac{2-p}{p-1}\right)\left(\frac{m_p}{m_e}\right)\epsilon_e
\gamma\gamma_M^{p-2}\right]^{1/(p-1)}.
\end{equation}
Equation (1) is different from the frequently-used minimum Lorentz factor 
$\gamma_m = [(p-2)/(p-1)](m_p/m_e)\epsilon_e\gamma$, 
which is also derived from these integrals for $p>2$. 
Consequently, we will obtain expressions for afterglow light curves  
that differ from derived by Sari et al. (1998, 1999). In addition, the age of 
the jet could also provide a limit on $\gamma_m$ through constraining 
$\gamma_M$ in equation (1). However, we have found that the value of 
$\gamma_M$ inferred from the age limit is usually much larger than the one 
from the limit that the acceleration time equals the synchrotron cooling time, 
and thus the age limit on $\gamma_m$ can be ignored. The $\gamma_c$, 
the Lorentz factor of electrons that cool on the expansion time, is given by
$\gamma_c=6\pi m_ec/(\sigma_T\gamma B^2t)$, where $t$ is the observer's 
time (neglecting the redshift correction) (Sari et al. 1998). After having 
$\gamma_m$ and $\gamma_c$, we can easily obtain the evolution of $\nu_m$ 
and $\nu_c$ with time based on $\nu_m\propto \gamma\gamma_m^2B$ 
and $\nu_c\propto \gamma\gamma_c^2B$ (see below). The remaining 
break frequency is the self-absorption one, which is given by
$\nu_a=\{5enr/[(3-s)B\gamma_m^5]\}^{3/5}\nu_m$  
for $\nu_a\ll \nu_m<\nu_c$ (expected at late times of 
the afterglow, cf. Panaitescu \& Kumar 2000).

The next crucial question is how the Lorentz factor $\gamma$ decays 
with the observer's time because the break frequencies and the peak
flux, which are needed in calculating the observed flux, are functions 
of $\gamma$ and of the shock radius and medium density. Even if 
the shock is beamed, as long as $\gamma>\theta_0^{-1}(c_s/c)$, 
the jet evolution is a sphere-like expansion based on the 
Blandford-McKee (1976) self-similar solution, and thus the Lorentz 
factor decreases as $\gamma=8.2(E_{53}/n_*)^{1/8}t^{-3/8}$ for $s=0$, 
or $\gamma=8.8(E_{53}/A_*)^{1/4}t^{-1/4}$ for $s=2$,
where $E_{53}$ is the isotropic-equivalent energy of the jet in units of
$10^{53}$ ergs, and $t$ is in units of 1 day. However, the transition of 
the jet evolution takes place at $\gamma\sim \theta_0^{-1}(c_s/c)$, 
which in fact defines the break time $t_b$.
After this time, the jet will enter the spreading phase with 
$c_s=c/\sqrt{3}$ in Rhoads (1999) or with $c_s=c$ in Sari et al. (1999).
As a result, the Lorentz factor decays as $\gamma\propto t^{-1/2}$. 

After knowing the evolution of $\gamma$, we can find scaling 
relations of the break frequencies with time.  First,   
we derive the characteristic frequency 
\begin{equation}
\nu_m \propto \left \{
       \begin{array}{lll}
         t^{-[3(p+2)]/[8(p-1)]}, & {\rm spherical\,\, in\,\,ISM},\\   
         t^{-(p+4)/[4(p-1)]},   & {\rm spherical\,\, in\,\,wind},\\
         t^{-(p+2)/[2(p-1)]},          & {\rm jet}.
        \end{array}
       \right.
\end{equation}
Second, the self-absorption frequency
is found to evolve as 
\begin{equation}
\nu_a \propto \left \{
       \begin{array}{lll}
         t^{[9(2-p)]/[16(p-1)]}, & {\rm spherical\,\,in\,\,ISM},\\   
         t^{(74-49p)/[40(p-1)]},   & {\rm spherical\,\,in\,\,wind},\\
         t^{(34-19p)/[20(p-1)]},          & {\rm jet}.
        \end{array}
       \right.
\end{equation}
Finally, the cooling frequency evolves as $\nu_c \propto t^{-1/2}$ for 
a spherical shock in the ISM,  $\nu_c \propto t^{1/2}$ for a spherical shock 
in the wind, and $\nu_c \propto t^{0}$ for a jet.
        
In addition, the observed peak flux, $F_{\nu_m}$, has been derived 
by many authors (e.g., Waxman 1997;  Dai \& Lu 1998; Wijers \& 
Galama 1999; Rhoads 1999; Sari et al. 1998, 1999; Chevalier \& Li 2000). 
In this Letter we neglect the effect of dust extinction 
on the peak flux because this effect has been discussed to be significant 
only for a highly collimated jet expanding in a dense circumstellar cloud 
by Dai, Huang \& Lu (2001).  

Therefore, we can calculate the light curves for four frequency ranges.
The flux at frequencies lower than $\nu_a$:  $F_{\nu<\nu_a}
= F_{\nu_m}(\nu_a/\nu_m)^{1/3}(\nu/\nu_a)^2$, and thus evolves as 
\begin{equation}
F_{\nu<\nu_a}\propto \left \{
       \begin{array}{lll}
         t^{(17p-26)/[16(p-1)]}, & {\rm spherical\,\,in\,\,ISM},\\   
         t^{(13p-18)/[8(p-1)]},   & {\rm spherical\,\, in\,\,wind},\\
         t^{[3(p-2)]/[4(p-1)]},          & {\rm jet}.
        \end{array}
       \right.
\end{equation}
The flux above the self-absorption frequency but below 
the characteristic frequency is given by $F_{\nu_a<\nu<\nu_m}=
F_{\nu_m}(\nu/\nu_m)^{1/3}$, which evolves as
\begin{equation}
F_{\nu_a<\nu<\nu_m}\propto \left \{
       \begin{array}{lll}
         t^{(p+2)/[8(p-1)]}, & {\rm spherical\,\,in\,\,ISM},\\   
         t^{[5(2-p)]/[12(p-1)]}, & {\rm spherical\,\, in\,\,wind},\\
         t^{(8-5p)/[6(p-1)]},  & {\rm jet}.
        \end{array}
       \right.
\end{equation}
It is seen from equations (4) and (5) that, below $\nu_m$, 
the light curve index is still determined by $p$. As a comparison, 
the index is independent of $p$ in the case of $p>2$ (Sari et al. 1999). 
For $p=1.5$ (similar to the index obtained by Malkov 1999 for Fermi 
acceleration in the limit when particles acquire a significant fraction 
of the shock energy), the flux at $\nu<\nu_a$ is approximately 
constant, and the flux at $\nu_a<\nu<\nu_m$ increases as 
$\propto t^{7/8}$ for the ISM case and $\propto t^{5/12}$ for 
the wind case, respectively, as long as the expansion is spherical. Then, once 
the jet enters the spreading phase, the flux below the self-absorption 
frequency begins to decline as $\propto t^{-0.75}$, and the flux at higher 
frequency begins to increase slowly as $\propto t^{1/6}$. 

If the observed high-frequency emission comes from the  
radiating electrons that are slow cooling, we have its flux
$F_{\nu_m<\nu<\nu_c}=F_{\nu_m}(\nu/\nu_m)^{-(p-1)/2}$, which decays as
\begin{equation}
F_{\nu_m<\nu<\nu_c} \propto  \left \{
       \begin{array}{lll}
         t^{-3(p+2)/16}, & {\rm spherical\,\, in\,\,ISM},\\   
         t^{-(p+8)/8},   & {\rm spherical\,\, in\,\,wind},\\
         t^{-(p+6)/4},          & {\rm jet}.
        \end{array}
       \right.
\end{equation}
Above the cooling frequency, we obtain $F_{\nu>\nu_c} = 
F_{\nu_m}(\nu_c/\nu_m)^{-(p-1)/2}(\nu/\nu_c)^{-p/2}$, which declines as 
\begin{equation}
F_{\nu>\nu_c} \propto \left \{
       \begin{array}{lll}
         t^{-(3p+10)/16}, & {\rm spherical\,\, in\,\,ISM},\\   
         t^{-(p+6)/8},   & {\rm spherical\,\, in\,\,wind},\\
         t^{-(p+6/4},          & {\rm jet}.
        \end{array}
       \right.
\end{equation}
Bhattacharya (2001) derived light curves of the emission from a jet expanding 
in the ISM by assuming a general case of $\gamma_M\propto \gamma^q$ in 
equation (1). Our light curves in equations (6) and (7) are consistent with his 
$q=-1/2$ result. Define the light curve index $\alpha$ and the spectral index 
$\beta$ through $F_\nu (t)\propto t^{-\alpha}\nu^{-\beta}$. Table 1 summarizes 
the relations between $\alpha$ and $\beta$ above $\nu_m$ for different cases. 
Figures 1 and 2 further present the $\alpha-\beta$ relations 
for $1<p<2$ as well as those for $p>2$, in the ISM and wind cases, 
respectively. We see that, for each line in these figures,
the $p<2$ segment is not an extrapolation of the $p>2$ segment. 

\begin{center}
\begin{table*}[ht!]
\begin{center}
\begin{tabular}{|c||c||c|c|c|}
\hline
& spectral index $\beta$ & \multicolumn{3}{|c|}{light curve index $\alpha$
($F_{\nu}\propto t^{-\alpha}$)} \\ 
frequency & ($F_{\nu}\propto \nu^{-\beta}$) & sphere in ISM & 
sphere in wind & jet \\ \hline\hline
&  & $\alpha=3(p+2)/16$ & $\alpha=(p+8)/8$ & $(p+6)/4$ \\ 
\raisebox{1.5ex}[0pt]{$\nu<\nu_c$} & \raisebox{1.5ex}[0pt]{$\beta=(p-1)/2$}
& $\alpha=3(2\beta+3)/16$ & $\alpha=(2\beta+9)/8$ & 
$(2\beta+7)/4$ \\ \hline
&  & $\alpha=(3p+10)/16$ & $\alpha=(p+6)/8$ & $(p+6)/4$ \\ 
\raisebox{1.5ex}[0pt]{$\nu>\nu_c$} & \raisebox{1.5ex}[0pt]{$\beta=p/2$}
& $\alpha=(3\beta+5)/8$ & $\alpha=(\beta+3)/4$ & 
$(\beta+3)/2$ \\ \hline
\end{tabular}
\end{center}
\par
\label{t:afterglow}
\caption{The spectral index $\beta$ and the light curve index $\alpha$ as
function of $p$ in the case of $1<p<2$.   The parameter-free 
relation between $\alpha$ and $\beta$ is given for each case 
by eliminating $p$.}
\end{table*}
\end{center}

\section{Comparison with the Afterglow of GRB 010222}

The $UBVRI$ light curve of the GRB 010222 afterglow has been fitted 
by one broken power law: $F_\nu\propto t^{-\alpha_1}$ before the break 
time $t_b$ and $F_\nu\propto t^{-\alpha_2}$ after $t_b$. Here we summarize 
the light curve indices, the break time, and the spectral index given in the literature: 
($\alpha_1$, $\alpha_2$, $t_b$, $\beta$)  are 
($0.60\pm 0.03$, $1.31\pm 0.03$, $0.48\pm 0.02$ days, $1.1\pm 0.1$) 
(Masetti et al. 2001),
($0.80\pm 0.05$, $1.30\pm 0.05$, $0.72\pm 0.10$ days, $0.88\pm 0.10$) 
(Stanek et al. 2001), and 
($0.74\pm 0.05$, $1.35\pm 0.04$, $0.7\pm 0.07$ days, $0.75\pm 0.02$) 
(Sagar et al. 2001b).
In addition, the X-ray decay index 
after the break measured by BeppoSAX is $\alpha_2=1.33\pm 0.04$ 
and the spectral index $\beta=0.97\pm 0.05$ (in 't Zand et al. 2001). 
A common result of the optical and X-ray observations is that 
the light curve indeed began to steepen to $\propto t^{-1.3}$ 
about 0.5 days after the GRB. This is the earliest observed 
break of all the studied afterglows.   

The temporal property of the afterglow from GRB 010222 
is naturally reminiscent of the jet model. Indeed, some authors 
(e.g., Stanek et al. 2001; Sagar et al. 2001b; Cowsik et al. 2001) attributed 
this afterglow to a highly collimated jet. Stanek et al. gave 
a spectral fit of their $BVRI$ data, and obtained an index of $\beta=0.88
\pm 0.10$, in excellent agreement with the $g'r'i'z'$ fit of Lee et al. (2001), 
$\beta=0.90\pm 0.03$, and with the spectral index given by Jha et al. (2001), 
$\beta=0.89\pm 0.03$. This implies a spectral index of the electron distribution, 
$p=2.8$ in the slow-cooling electron regime or $p=1.8$ in the fast-cooling electron 
regime. The former value of $p$ leads to $F_\nu\propto t^{-2.8}$ at late times 
(Rhoads 1999; Sari et al. 1999) while the latter value gives $F_\nu\propto t^{-1.95}$
(see section 2). These results are inconsistent 
with the observed late-time light curve ($\propto t^{-1.3}$).
Stanek et al. have noted this inconsistency. To save the jet model, they 
suggested that the spectral index could be intrinsically in the range of
$0.5<\beta < 0.7$ due to the SMC-like extinction. Such a range of the spectral 
index requires $2<p<2.4$ (slow cooling) or $1<p<1.4$ (fast cooling). Even if
the value of $p$ becomes smaller for the theoretical spectral index 
to be compatible with the observed extinction-corrected spectral 
index, according to Rhoads (1999), Sari et al. (1999) and our analysis 
in section 2, we still conclude that the spreading jet model cannot provide 
an explanation for the late-time light curve index. We note that 
Sagar et al. (2001b) suggested the afterglow of GRB 010222 as evidence
for a highly collimated jet with a fast-cooling, flat-spectrum electron 
distribution. Their argument is that the emission flux from a spreading jet 
decays as $\propto t^{-p}$ for $1<p<2$, which means  
$F_\nu\propto t^{-1.3}$ when $p=1.3$, inferred by their fitting spectrum. 
However, from our analysis in section 2, 
we see that their argument is incorrect.     

An alternative explanation for the afterglow of GRB 010222 is the expansion 
of a relativistic fireball or a mildly collimated jet  in a medium with density of 
$10^5-10^6\,\,{\rm cm}^{-3}$ (Masetti et al. 2001; in 't Zand et al. 2001). In such 
a dense medium, the fireball decelerated to the non-relativistic regime
within a few days after the burst, resulting in a steepening of the light curve
(Dai \& Lu 1999, 2000). in 't Zand et al. argued that the non-relativistic 
interpretation {\em with a universal $p\approx 2.2$ value} is consistent 
with the observations. They also noted that the dense-medium assumption 
is compatible with the observed redshift-corrected column density of 
$\sim 2.5\times 10^{22}\,\,{\rm cm}^{-2}$.  

\section{Discussion and Conclusion}      

We have derived light curves of the emission when 
a highly collimated jet with a flat-spectrum electron distribution 
($1<p<2$) expands in the ISM or in the pre-burst wind. The most 
important finding of ours is that once the jet begins to spread, the light curve 
index becomes $(p+6)/4$ rather than $p$. Therefore, the jet model 
appears to be inconsistent with the afterglow data of GRB 010222.  
ICS in the shocked medium doesn't influence the light curves derived 
in section 2. This is because for $p<2$ most of the electron energy behind 
the shock front should be radiated away via both synchrotron radiation and ICS
and thus the Compton parameter $Y\approx (-1+\sqrt{1+4\epsilon_e/\epsilon_B})/2$ 
is approximately constant (Panaitescu \& Kumar 2000; Sari \& Esin 2001).  

Two important quantities that future observations led by {\em HETE-2} 
and {\em Swift} will provide are the light curve index and the spectral index, 
which, once known, will show a point in Figures 1 and 2. According to 
the position of this point in these figures, one could not only obtain information 
on the dynamical evolution of a post-burst shock wave and the radiation 
regime of the accelerated electrons (slow cooling or fast cooling), 
but also infer the value of $p$. 

It should be emphasized that our derivations in section 2 
are based on the assumption that the electron energy density behind a shock 
is a constant fraction ($\epsilon_e$) of the total energy density of the shocked 
medium, as used in the standard afterglow shock model. If this assumption 
is invalid, the minimum Lorentz factor of the electrons, without any 
acceleration, could become $\gamma$ instead of equation (1). In such a case, 
the previous jet model could explain the afterglow of GRB 010222 
if the electron energy distribution is required to be a power law with $p<2$.
However, it is unclear whether this requirement is satisfied in 
the absence of any acceleration.          
      
\acknowledgments
We are very grateful to the referee and D. M. Wei for valuable comments 
that significantly improved the manuscript,  and to N. Masetti and B. Zhang 
for discussions. This work was supported by a RGC grant of Hong Kong 
government, the National Natural Science Foundation of China 
(grant 19825109), and the National 973 Project (NKBRSF G19990754).

\clearpage
\begin{figure}
\begin{picture}(100,250)
\put(0,0){\includegraphics{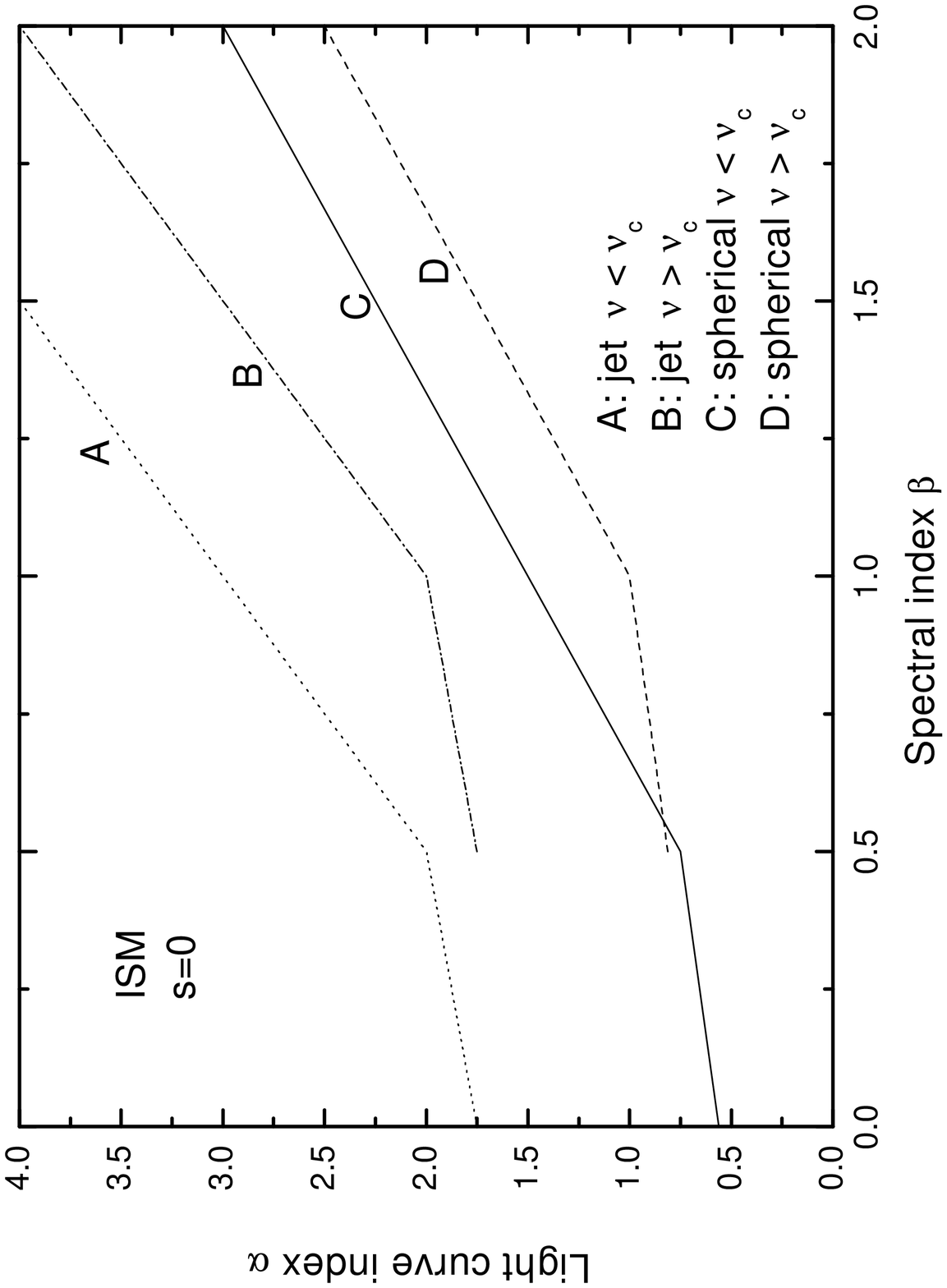}}
\end{picture}
\caption
{A plot of the light curve index ($\alpha$) versus the spectral index 
($\beta$) in the ISM case. Lines A and B correspond to a highly 
collimated but spreading jet whose (observed) high-frequency emission 
comes from the radiating electrons that are slow cooling ($\nu<\nu_c$) 
and fast cooling ($\nu>\nu_c$), respectively, and lines C and D to 
a spherical shock. }
\end{figure}

\clearpage
\begin{figure}
\begin{picture}(100,250)
\put(0,0){\includegraphics{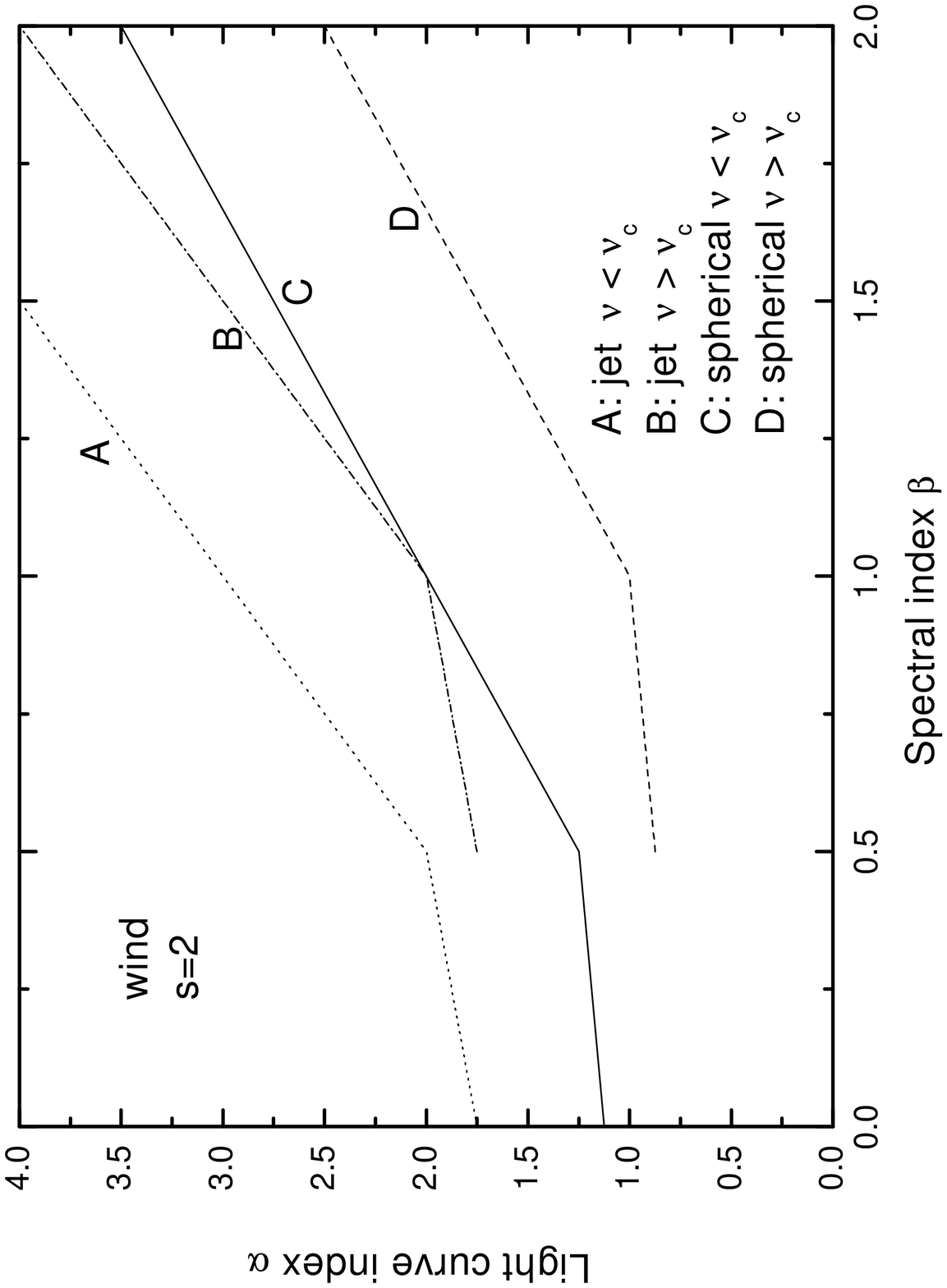}}
\end{picture}
\caption
{Same as Fig. 1 but in the wind case.}  
\end{figure}

\end{document}